\documentclass[12pt]{iopart}
\usepackage{iopams,setstack,graphicx,color,bm} 
\usepackage[colorlinks, citecolor=green, linkcolor=blue]{hyperref}

\definecolor{blue4}{rgb}{0.000, 0.000, 0.508}
\definecolor{red3}{rgb}{0.807, 0.000, 0.000}
\definecolor{orange}{rgb}{1.000, 0.647, 0.000}

\newcommand{\Orightarrow}[1]{\overset{{#1}}{\longrightarrow}}

\newcommand{\Urightarrow}[1]{\underset{{#1}}{\longrightarrow}}
\newcommand{\Uleftarrow}[1]{\underset{{#1}}{\longleftarrow}}
\newcommand{\E}[2]{\mathbb{E}_{#1}\left[ #2 \right]}

\begin{document}

\title[]{Can we always sweep the details of RNA-processing under the carpet?}

\author{Filippos D. Klironomos$^1$, Juliette de Meaux$^{2}$, and Johannes Berg$^{1,3}$}
\address{$^1$ Institute for Theoretical Physics, University of Cologne, Cologne, Germany}
\address{$^2$ Institute for Evolution and Biodiversity, University of M\"unster; M\"unster, Germany}
\address{$^3$ Systems Biology of Ageing - Sybacol, Cologne, Germany}
\ead{fklirono@uni-koeln.de}

\begin{abstract}

RNA molecules follow a succession of enzyme-mediated processing steps from transcription until maturation.
The participating enzymes, for example the spliceosome for mRNAs and Drosha and Dicer for microRNAs,
are also produced in the cell and their copy-numbers fluctuate over time.
Enzyme copy-number changes affect the processing rate of the substrate
molecules; high enzyme numbers increase the processing probability, low enzyme numbers decrease it.
We study different RNA processing cascades where enzyme copy-numbers are either fixed or fluctuate.
We find that for fixed enzyme-copy numbers the substrates at steady-state are Poisson-distributed, 
and the whole RNA cascade dynamics can be understood as a single
birth-death process of the mature RNA product. In this case, solely 
fluctuations in the timing of RNA processing lead to variation  in the
number of RNA molecules. 
However, we show analytically and numerically that when enzyme copy-numbers fluctuate, the strength of RNA fluctuations 
increases linearly with the RNA transcription rate. 
This linear effect becomes stronger as the speed of enzyme dynamics decreases relative to the speed of RNA dynamics.
Interestingly, we find that under certain conditions, the RNA cascade can reduce the strength of fluctuations in the expression level of the mature RNA product.
Finally, by investigating the effects of processing polymorphisms we show that it is possible for the effects of transcriptional polymorphisms
to be enhanced, reduced, or even reversed.
Our results provide a framework to understand the dynamics of RNA processing.

\end{abstract}


\maketitle

\section{Introduction}

The copy-number of enzymes that mediate particular reactions is a source of intrinsic fluctuations in gene expression~\cite{Swain2002,Paulsson2005}.
An enzyme chemically converts its substrate at a rate which may be either fixed, or may change over time.
However, even for a fixed rate of chemical conversions per enzyme, prolonged changes in the enzyme copy-number become noticeable at the level of the enzyme's substrate.
Consequently, the substrates convert at a rate that fluctuates over time.
In this work, we consider the effect of such fluctuations on the production and processing of mRNA, microRNA (miRNA) or small interfering RNA (siRNA). 

The principal enzyme involved in mRNA-processing is the spliceosome, which removes intronic elements from precursor mRNA molecules~\cite{Wahl2009,Will2011}.
In the biogenesis of eukaryotic small RNAs (sRNA), encompassing miRNAs and siRNAs, one additionally finds
(i) the microprocessor unit responsible for processing primary miRNAs, 
(ii) the RNA-dependent RNA polymerase responsible for synthesizing complementary strands to single-stranded RNA,
and (iii) the nuclease Dicer responsible for processing precursor sRNA molecules~\cite{Krol2010}.
In the biogenesis of prokaryotic sRNA one finds the Cascade/Cas enzymes performing similar processing steps~\cite{Vanderoost2009}.
The processed 
eukaryotic or prokaryotic 
sRNA molecules are then loaded to Argonaute proteins forming the so called RNA-induced silencing complexes (RISCs),
the units responsible for the post-transcriptional regulation (PTR) of
mRNA transcripts~\cite{Rhoades2006,Kim2009}.

The enzymes involved in the mRNA and sRNA biogeneses are vital for the normal functioning of the cell and are often regulated by complex feedback networks.
For example, miR162 targets the Dicer DCL1 mRNA in \textit{Arabidopsis thaliana}, but the precursor of miR162 needs Dicer to mature~\cite{Allen2005}.
Hence large expression levels of Dicer lead to large levels of the miRNA that in turn downregulates Dicer.
However, the presence of tight regulation of the enzymatic expression levels does not eliminate completely fluctuations in their copy-numbers.
At best, regulation provides a mechanism to limit the range of fluctuations around the mean value of enzymatic copy-numbers at steady-state.

Studies of the protein biogenesis usually assume mRNA is produced via a birth-death process~\cite{Thattai2001,Shahrezaei2008}.
Studies of post-transcriptional regulation involving sRNAs assume the same~\cite{Shimoni2007,Levine2008,Hao2011,Wang2011,Klironomos2013}.
This assumption that the multi-step process of RNA biogenesis can be modeled by a single birth-death process of the mature RNA product requires examination.
We address this question by studying the mRNA and sRNA biogeneses using models that explicitly include the sequential processing of RNA precursors by the different enzymes.
We find that under certain conditions one can indeed replace the complex RNA-processing cascade by a single production process of constant rate. 
In that case, mature RNA production events are statistically independent and follow a homogeneous Poisson process.
However, outside the validity of these conditions, this simple picture breaks down since fluctuations in a RNA processing cascade cannot be captured by 
a single homogeneous Poisson process.
In this case, enzymatic copy-number fluctuations introduce statistical dependencies in the RNA-processing events.

\section{Results}

We start with the simplest scenario possible, a chain of RNA-processing steps.
Such RNA-processing cascades arise in different contexts.

\subsection{mRNA biogenesis}
\label{mRNA}

The biogenesis of messenger RNA starts with the transcription from DNA of the primary RNA transcript (pre-mRNA$_n$),
and continues as the spliceosome excises introns until the final mRNA product is reached.
If we assume each of these steps takes place at a constant rate, the cascade of events can be depicted schematically as follows
\begin{equation}
\def\arraystretch{0.8}  %  reduce row spacing
\begin{array}{ccccc}
\varnothing & \Orightarrow{k_n} &  \mathrm{pre\!\!-\!\!mRNA}_n  & \Orightarrow{d_n} & \varnothing, \\ 
 \mathrm{pre\!\!-\!\!mRNA}_n  & \Orightarrow{k_{n-1}} &  \mathrm{pre\!\!-\!\!mRNA}_{n-1}  & \Orightarrow{d_{n-1}} & \varnothing, \\
\vdots  & \vdots & \vdots & \vdots & \\
 \mathrm{pre\!\!-\!\!mRNA}_1  & \Orightarrow{k_{0}} &  \mathrm{mRNA}  & \Orightarrow{d_0} & \varnothing,
\end{array}  \label{R:mRNA}
\end{equation}
where $k_n$ is the gene transcription rate, $k_{n-1},\ldots,k_0$ the precursor mRNA processing rates,
and $d_n,\ldots,d_0$ the basal degradation rates of all mRNA products.
The length of the cascade is determined by the number of introns $n$ of the primary transcript.
We will show that at steady-state and under certain conditions, each of the components of the mRNA
cascade is Poisson-distributed, and the mRNA creation dynamics follows a homogeneous Poisson process.
In that case, (\ref{R:mRNA}) can be replaced 
by $\varnothing \Orightarrow{\widetilde{k}_0} \mathrm{mRNA} \Orightarrow{d_0} \varnothing$
if $\widetilde{k}_0$ is chosen to match the effective production rate of mRNA at steady-state in (\ref{R:mRNA}).

\subsection{siRNA biogenesis}
\label{siRNA}

A second example of a chain of RNA processing is the 
biogenesis of endogenous small interfering RNAs (siRNA).
This process starts with the transcription of genes or transposable elements to single-stranded RNA (ssRNA).
The ssRNA is then converted by RNA-dependent RNA polymerase (RDR) to double-stranded RNA (dsRNA), which is in turn 
cleaved by Dicer nuclease to mature siRNAs.
The biogenesis of exogenous siRNAs, involved for example in infections or transfections,
is delocalized with only the last processing step occurring at the place where the siRNAs operate.
In any case, if we assume that each step takes place at constant rate, then 
the siRNA cascade can be depicted as follows
\begin{equation}
\def\arraystretch{0.8}  %  reduce row spacing
\begin{array}{ccccc}
\varnothing & \Orightarrow{k_2} &  \mathrm{ssRNA}  & \Orightarrow{d_2} & \varnothing, \\ 
 \mathrm{ssRNA}  & \Orightarrow{k_1} &  \mathrm{dsRNA}  & \Orightarrow{d_1} & \varnothing, \\
 \mathrm{dsRNA}  & \Orightarrow{k_0} &  \mathrm{siRNA}  & \Orightarrow{d_0} & \varnothing.
\end{array}  \label{R:siRNA}
\end{equation}
This reaction scheme is analogous to the mRNA-processing chain (\ref{R:mRNA}) and
can be simplified under certain conditions to a single birth-death process as discussed below.

Not included in this scheme is the biogenesis of trans-acting small interfering RNA (tasiRNA).
For this particular class of siRNAs, the assumption of constant rate of production $k_2$ of ssRNA in (\ref{R:siRNA}) is invalid:
the tasiRNA biogenesis initiates when fragments of post-transcriptionally regulated mRNAs, cleaved with the help of miRNAs, 
are converted to dsRNA by RDR~\cite{Rajagopalan2006,Chen2010}.
This process can fluctuate considerably resulting in temporal changes in the production rate $k_2$ of ssRNA~\cite{Klironomos2013}.

\subsection{miRNA biogenesis}
\label{miRNA}

MiRNAs originate from either intergenic DNA regions having their own promoters, or intragenic DNA regions of introns of protein-encoding genes~\cite{Krol2010}. 
The latter can vary in length: short intragenic fragments called mirtrons are processed directly by the spliceosome~\cite{Westholm2011}, 
whereas longer intragenic fragments are additionally processed by the microprocessor unit in animals~\cite{Zhu2008},
and by Dicer in plants~\cite{Zhu2008,Rajagopalan2006}.
There is evidence that the processing order might matter, and also that the microprocessor unit and the spliceosome collaborate in their tasks~\cite{Janas2011}.
The latter suggests that these two processing steps might not be entirely independent.
When the microprocessor unit (or Dicer in plants) acts first, the intragenic miRNA biogenesis can be depicted as follows
\begin{equation}
\def\arraystretch{0.8}  %  reduce row spacing
\begin{array}{ccccc}
\varnothing & \Orightarrow{k_2} &  \mathrm{pri\!\!-\!\!miRNA_{pre-mRNA}}  & \Orightarrow{d_2} & \varnothing, \\ 
 \mathrm{pri\!\!-\!\!miRNA_{pre-mRNA}}  & \Orightarrow{k_1} &  \mathrm{pre\!\!-\!\!miRNA}  & \Orightarrow{d_1} & \varnothing, \\
 \mathrm{pre\!\!-\!\!miRNA}  & \Orightarrow{k_0} &  \mathrm{miRNA}  & \Orightarrow{d_0} & \varnothing,
\end{array}  \label{R:miRNA-intra}
\end{equation}
where pri-miRNA$_{\mathrm{pre-mRNA}}$ denotes the long RNA hairpin consisting of the primary miRNA intragenic element and the precursor mRNA fragment.
If the spliceosome acts first on the long RNA hairpin it produces pri-miRNA from the intronic fragment, 
which is consequently processed into pre-miRNA by the microprocessor unit (or Dicer in plants). 
Mirtrons have a biogenesis identical to (\ref{R:miRNA-intra}), but
have shorter pri-miRNAs and the processing of the long primary RNA hairpin is performed by the spliceosome only~\cite{Westholm2011}.

Intergenic miRNA follows a similar biogenesis as intragenic miRNA:
\begin{equation}
\def\arraystretch{0.8}  %  reduce row spacing
\begin{array}{ccccc}
\varnothing & \Orightarrow{k_2} &  \mathrm{pri\!\!-\!\!miRNA}  & \Orightarrow{d_2} & \varnothing, \\ 
 \mathrm{pri\!\!-\!\!miRNA}  & \Orightarrow{k_1} &  \mathrm{pre\!\!-\!\!miRNA}  & \Orightarrow{d_1} & \varnothing, \\
 \mathrm{pre\!\!-\!\!miRNA}  & \Orightarrow{k_0} &  \mathrm{miRNA}  & \Orightarrow{d_0} & \varnothing.
\end{array}  \label{R:miRNA-inter}
\end{equation}
The only difference between intergenic and intragenic miRNA biogenesis
is the additional production of mRNA for the latter after the spliceosome
excises all introns from the pri-miRNA$_{\mathrm{pre-mRNA}}$ hairpin.
Both (\ref{R:miRNA-intra}) and (\ref{R:miRNA-inter}) have a structure
similar to the mRNA cascade (\ref{R:mRNA}).
When reaction rates are constant, we show below that the miRNA cascade can be replaced by 
$\varnothing \Orightarrow{\widetilde{k}_0}  \mathrm{miRNA}  \Orightarrow{d_0} \varnothing$.
If the host transcript in the intragenic miRNA biogenesis is also a member of the network investigated,
then it is straightforward to include its biogenesis separately.

A complication outside the applicability of (\ref{R:miRNA-intra})
arises if the mature intragenic miRNA regulates the host transcript post-transcriptionally.
The majority of intragenic miRNAs investigated across species (80\%) are predicted not to target their hosts~\cite{Hinske2010}.
On the other hand, the biogenesis of the remaining 20\% couples the production rates of miRNA and target, 
and may force PTR to operate close to the so called ``derepression threshold"~\cite{Levine2008,Mukherji2011},
where targets are expressed at levels that are just sufficient to overcome repression via PTR.
In this regime of target expression and beyond, the RISC-formation and RISC-recycling processes play a prominent role:
they control the strength of PTR-induced fluctuations in the target transcript levels~\cite{Klironomos2013}.
Finally, we should mention that post-transcriptional regulation and feedback loops~\cite{Gjuvsland2007,Wang2011} are not within the scope of this work.
We focus solely on RNA-processing so that we can understand its underlying dynamics.

\subsection{Unifying the RNA cascades under constant reaction rates}

The mRNA (\ref{R:mRNA}) and sRNA (\ref{R:siRNA}-\ref{R:miRNA-inter}) biogeneses under constant reaction rates 
are all special cases of the following generic cascade
\begin{equation}
\def\arraystretch{0.5}  %  reduce row spacing
\begin{array}{ccccc}
\varnothing & \Orightarrow{k_x} &  x & \begin{array}{c} \Orightarrow{k_y} \\ \Urightarrow{d_x} \end{array} & \varnothing, \\
 x & \Orightarrow{k_y} &  y & \begin{array}{c} \Orightarrow{k_z} \\ \Urightarrow{d_y} \end{array} & \varnothing, \\
 y & \Orightarrow{k_z} &  z & \Orightarrow{d_z} & \varnothing.
\end{array}  \label{R:RNA}
\end{equation}
Here, the processing steps are broken down into two sub-steps:
one step involving the destruction of the ancestor precursor, and one step involving the creation of the new product.
The reason we choose this representation is because it separates processes according to intermediate components. 
However, both processing reactions in (\ref{R:RNA}) are not independent but rather take place simultaneously.
For example, the process $x \Orightarrow{k_y} \varnothing$ never takes place without the partner reaction $ x \Orightarrow{k_y}y$.

Below we show analytically that at steady-state the $x$, $y$, and $z$ products of (\ref{R:RNA}) are Poisson-distributed.
The rest of the details in (\ref{R:mRNA}-\ref{R:miRNA-inter}) not included in (\ref{R:RNA}), for example the number of introns excised in (\ref{R:mRNA})
influence the average steady-state expression level of the mature product, \textit{i.e.,} the mean value of the Poisson distribution.
All else being equal, the expression levels of two mature mRNAs for example, might be different if one has introns in its precursor and the other does not, 
but they are both going to be Poisson-distributed at steady-state if processing rates per precursor mRNA are constant.

\subsection{Solution of the RNA biogenesis under constant reaction rates}

We consider the simplest scenario in some detail here as the same tools will be used to treat the effects of fluctuating reaction rates in section~\ref{fluctuating_rates}.
The master equation describing the generic RNA biogenesis (\ref{R:RNA}) is
\begin{eqnarray}
\frac{d \rho_{ x,  y,  z}}{dt} &= 
\left[ k_x\left(\mathcal{E}^{-}_{ x} - 1\right) + d_x\left(\mathcal{E}^{+}_{ x}-1\right) x
+ k_y\left(\mathcal{E}^{+}_{ x}\mathcal{E}^{-}_{ y}-1\right) x \right. \nonumber\\
&\left. + d_y\left(\mathcal{E}^{+}_{ y} - 1\right) y + k_z\left(\mathcal{E}^{+}_{ y}\mathcal{E}^{-}_{ z}-1\right) y
+ d_z\left(\mathcal{E}^{+}_{ z}-1\right) z\right] \rho_{ x,  y,  z}, \label{ME}
\end{eqnarray}
where $\rho_{ x,  y,  z}(t)$ is the probability for the molecular numbers in the system at a given time to be $ x,  y,  z$. 
The shift operators $\mathcal{E}^{\pm}$ are defined by $\mathcal{E}^{\pm}_{ n}g( n)=g( n\pm 1)$, where $ n\in\{ x,  y, z\}$.
Using the generating function $f(r, s, q,t) = \sum_{ x, y, z} r^x s^y q^z \rho_{x,y,z}(t)$ in (\ref{ME}) we arrive at
\begin{eqnarray}
\fl
\frac{\partial f(r,s,q,t)}{\partial t} &=
k_x(r-1)f(r,s,q, t) + \left[d_x(1-r) + k_y(s-r)\right] \frac{\partial f(r,s,q,t)}{\partial r} \nonumber \\
&+ \left[d_y(1-s) + k_z(q-s)\right] \frac{\partial f(r,s,q,t)}{\partial s} + d_z (1-q) \frac{\partial f(r,s,q,t)}{\partial q}. \label{ME:f}
\end{eqnarray}
At steady-state the product of three generating functions of Poisson distributions $f(r,s,q) = e^{\langle x\rangle (r-1)}e^{\langle y\rangle (s-1)}e^{\langle z\rangle (q-1)}$
solves (\ref{ME:f}) when
\begin{eqnarray}
\langle x\rangle = \frac{k_x}{d_x+k_y},\qquad \langle y\rangle = \frac{k_y\langle x\rangle}{d_y+k_z}, \qquad \langle z\rangle = \frac{k_z\langle y\rangle}{d_z}. \label{ss}
\end{eqnarray}
Consequently, the steady-state distribution of $z$ in the cascade (\ref{R:RNA}) is identical 
to the steady-state distribution of $z$ in the single birth-death process $\varnothing \Orightarrow{\widetilde{k}_z} z \Orightarrow{d_z} \varnothing$, where
$\widetilde{k}_z = k_z \langle y\rangle$.
As expected for Poisson-distributed quantities, we find $\langle n^2 \rangle - \langle n \rangle^2 = \langle n \rangle$ for $n=\{x,y,z\}$, 
and computing the Fano factor $F_n$,  defined as the variance over the mean of the random variable $n$, we obtain $F_x = F_y = F_z = 1$.

So far we have proved that at steady-state the mature product of
(\ref{R:RNA}) follows Poisson statistics.
However, we have not shown that the statistics production events of $z$ at steady-state in (\ref{R:RNA}) 
is identical to the production statistics of $z$ for a constant effective rate $\widetilde{k}_z$.
This question will be addressed in section \ref{dynamics} after we investigate the case of RNA biogenesis with fluctuating processing rates.

\subsection{Synergistic or antagonistic effects of polymorphisms in sRNA biogenesis}

Variation across organisms in sRNA processing dynamics can be two-fold:
(i) the same precursor sRNA can be processed differently by the RNA-processing enzymes producing several isoforms of the mature sRNA~\cite{Krol2010},
or (ii) different precursor sRNAs from different loci or different alleles across species can have different processing rates 
but produce identical mature sRNAs~\cite{Xie2005,Li2012,deMeaux2008,Todesco2012}.
In the first case, the processing variation affects the efficiency of recruitment of the mature sRNA isoforms by the Argonaute proteins~\cite{Juvvuna2012},
and consequently the recycling rate of recruited mature sRNAs after they have catalyzed a transcript-targeting event~\cite{Rhoades2006,Haley2004}.
Both of these effects have been addressed elsewhere~\cite{Klironomos2013}.
Here, we investigate the case of differences in the processing rates between two precursor sRNAs that give rise to identical mature sRNA products.

We assume two miRNA alleles ($miR_{1,2}$) produce identical mature miRNA products,
but have differences in the transcription and processing rates of their respective pri-miRNAs.
In particular, we assume that $miR_1$ is transcribed at a faster rate than $miR_2$ ($k_{x_1} > k_{x_2}$) but the primary transcript of $miR_1$ (pri-miR$_1$)
is processed at a slower rate ($k_{y_1} < k_{y_2}$). 
Is it possible that the steady-state copy-number $z_1$ of the mature miRNA product of $miR_1$ to be less than $z_2$, 
the identical corresponding mature miRNA product of $miR_2$?
The ratio of the steady-state expression levels of (\ref{ss}) yields
\begin{equation}
\frac{z_1}{z_2} = \frac{k_{x_1}}{k_{x_2}} \frac{1 + d_x / k_{y_2}}{1 + d_x / k_{y_1}}. \label{z1-z2}
\end{equation}
As long as pri-miR$_1$ is processed at a lower rate than pri-miR$_2$ ($k_{y_1} < k_{y_2}$), 
and despite the fact that $miR_1$ is transcribed at a higher rate than $miR_2$ ($k_{x_1} > k_{x_2}$), 
the steady-state expression level of mat-miR$_2$ can still be higher ($z_1<z_2$).
Similar results are obtained if variation is present in the processing rates of the precursor miRNAs instead of the primary miRNAs.
In other words,
the effects of polymorphisms in the transcription of sRNAs can be reversed or enhanced with the appearance of polymorphisms in their processing steps.

\subsection{RNA biogenesis with fluctuating processing rates}
\label{fluctuating_rates}

So far we considered mRNA and sRNA biogeneses with constant reaction
rates and showed that at steady-state the mature RNA follows Poisson statistics.
Now we consider changes in the processing rates due to fluctuations in enzyme copy-numbers.
We work with the generic RNA cascade (\ref{R:RNA}).

We define at a given instant the number of molecules of the enzymes that process RNA transcripts at the first and second step in (\ref{R:RNA})
as $\alpha$ and $\beta$, respectively.
By taking into account variations in enzyme copy-numbers, the RNA biogenesis becomes
\begin{equation}
\def\arraystretch{0.5}  %  reduce row spacing
\begin{array}{ccccc}
\varnothing & \Orightarrow{k_x} &  x & \begin{array}{c} \Orightarrow{k_y\alpha} \\ \Urightarrow{d_x} \end{array} & \varnothing, \\
 x & \Orightarrow{k_y\alpha} &  y & \begin{array}{c} \Orightarrow{k_z\beta} \\ \Urightarrow{d_y} \end{array} & \varnothing, \\
 y & \Orightarrow{k_z\beta} &  z & \Orightarrow{d_z} & \varnothing,
\end{array}  \label{RR:RNA}
\end{equation}
where the replacements $k_y \rightarrow k_y \alpha$ and $k_z \rightarrow k_z \beta$ are made in (\ref{R:RNA}).
That is, $k_y$ is now the constant conversion rate per $x$ molecule and per $\alpha$ enzyme. 
However, the processing rate $k_y \alpha(t)$ per $x$ molecule fluctuates now over time.
The same applies for the processing rate $k_z \beta(t)$ per $y$ molecule.

We denote as $1/\kappa$ the characteristic time scale over which enzyme expression level variation occurs.
If this characteristic time scale is much slower than the dynamics associated with the biogenesis
of RNA, then we can use the results of (\ref{ss}) for given values of $\alpha$, $\beta$ and ensemble-average over the equilibrium distributions of
$\alpha$, $\beta$.
The Fano factors $F_{n}$ for $n=\{x,y,z\}$ are given by
\begin{equation}
F_{n} = \frac{\E{\alpha,\beta}{\langle n^2\rangle} - \E{\alpha,\beta}{\langle n\rangle}^2}{\E{\alpha,\beta}{\langle n\rangle}} = 
1 + \frac{k_x}{d_x}\mathbb{F}[S_n], \label{F:n}
\end{equation}
where $\E{\alpha,\beta}{\cdot}$ indicates averaging over the equilibrium distributions of $\alpha$, $\beta$, 
and $\mathbb{F}[S_n]$ are Fano factors of the stochastic variables $S_n$ listed below
\begin{eqnarray}
\fl S_x  = \frac{1}{1 + k_y \alpha/d_x}, \;\;
S_y  = \frac{k_y \alpha/d_y}{1 + k_y \alpha/d_x} \frac{1}{1 + k_z \beta/d_y}, \;\;
S_z  = \frac{k_y \alpha/d_y}{1 + k_y \alpha/d_x} \frac{k_z\beta/d_z}{1 + k_z \beta/d_y}. \label{Sxyz}
\end{eqnarray}
No assumption has been made about the particular form of the distributions of $\alpha$, $\beta$.
The only assumptions are: (i) the distribution of enzyme expression levels is in steady-state,
and (ii) the enzyme biogenesis dynamics is slower than the RNA biogenesis dynamics, allowing us to employ the adiabatic approximation.
Formally, the latter condition is expressed as $\kappa \ll \max\{k_y \alpha, k_z \beta, d_x, d_y, d_z\}$.

How do fluctuations in $\alpha$, $\beta$ introduce a linear dependence on $k_x$ in (\ref{F:n})?
The answer lies in the way enzyme copy-number fluctuations imprint on substrates.
Any change in $\alpha$ or $\beta$ during time intervals that are similar or longer than the typical time interval of substrate biogenesis
is felt by the corresponding substrate.
The strength of this change is always related to the abundance of the corresponding substrate which in turn is proportional to $k_x$.
Therefore, the more abundant that substrate become, the larger that the effect of enzyme copy-number fluctuations becomes on them as well.
Furthermore, the strength of this effect depends also on the timescale of enzyme biogenesis. 
For example, faster enzyme copy-number fluctuations have less of an effect over the longer time scales of substrate biogenesis. 
In this case, processing of the substrate seems to be taking place under almost constant rates and the Fano factors of (\ref{F:n}) tend to unity.
On the other hand, slower fluctuations in the enzyme biogeneses induce slower changes in the processing rates, 
which are perceived by the substrates and render their biogeneses more noisy.

\begin{figure}[thb!]
\begin{center}
\includegraphics[totalheight=11.0cm, keepaspectratio=true, viewport=0 0 785 715,clip]{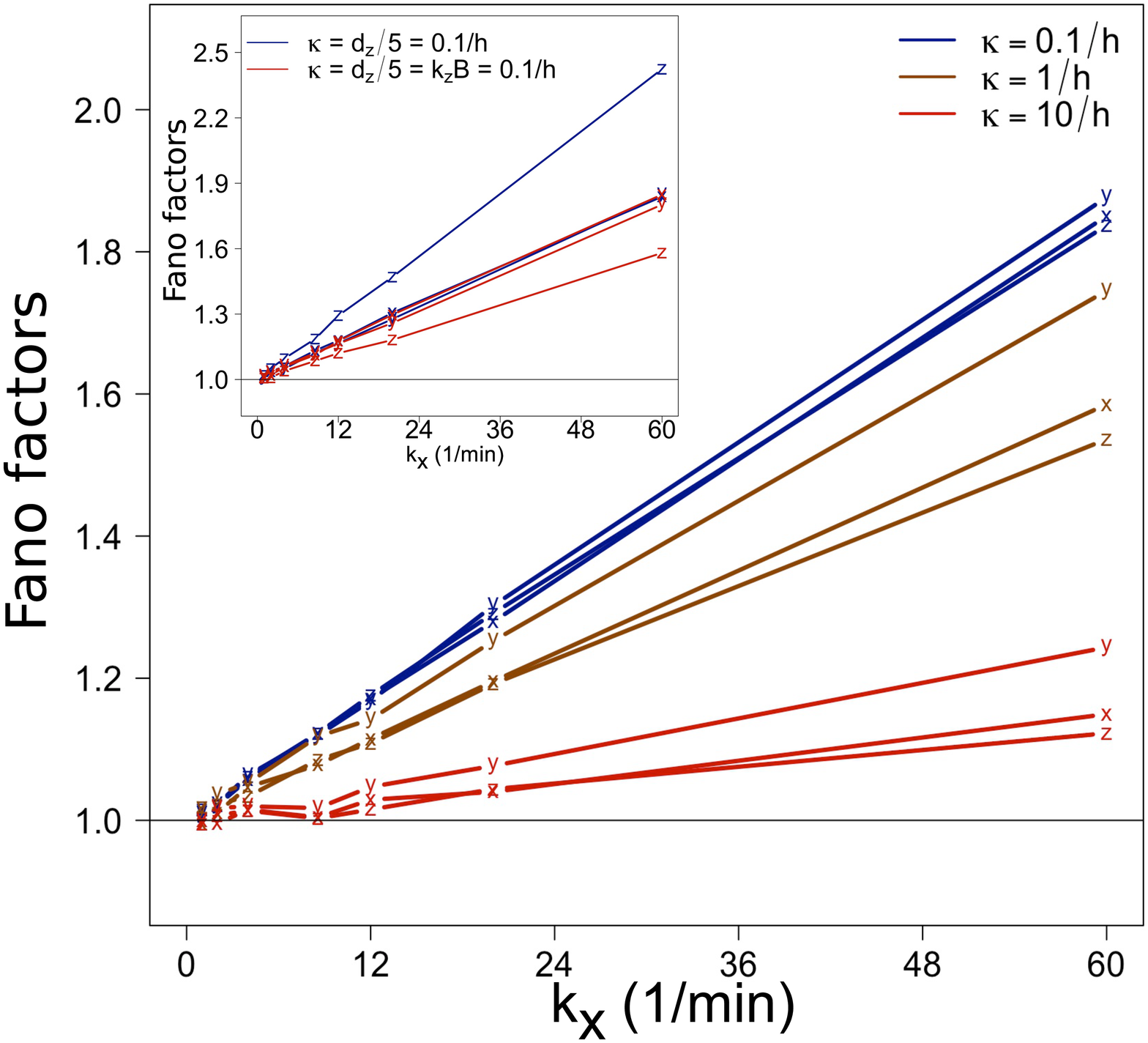}
\caption{\small
{\bf Strength of fluctuations in RNA biogenesis depend linearly on the RNA transcription rate.}
The Fano factors of $x$, $y$, and $z$ in (\ref{RR:RNA}) are plotted as functions of the transcription rate $k_{x}$.
The parameter values used are typical for the siRNA biogenesis in \textit{Salmonella}~\cite{Overgaard2009}:
$k_y A =k_z B=d_x=d_y=d_z=$1/h, $A=B=500$.
In the main plot the rate $\kappa$ of the biogeneses of the RNA-processing enzymes is varied.
Blue lines correspond to $\kappa=0.1$/h, brown lines correspond to $\kappa=1$/h, and red lines to $\kappa=10$/h.
The horizontal black line indicates the range of unity Fano factors predicted by (\ref{ME}) for Poissonian statistics.
Variation in enzyme copy-numbers induces fluctuations in the substrates, whose amplitude depends linearly on the RNA transcription rate.
For $\kappa \geq 0.1$/h and for $k_x\gg 1$/min we find $F_z < F_x < F_y$: the RNA cascade buffers the fluctuations in $z$.
However in the inset, we plot substrate Fano factors for twice as stable mature RNA ($d_z=0.5$/h) and for $\kappa=0.1$/h
and find $F_x \simeq F_y < F_z$ (blue lines): the RNA cascade amplifies the strength of fluctuations in $z$.
If on the other hand, the average processing rate of this stable mature RNA is also reduced ($k_z B=0.1$/h), the relation $F_x \simeq F_y > F_z$ is restored (red lines)
and the RNA cascade buffers again the fluctuations in $z$.
\label{fig:xyz}
}
\end{center}
\end{figure}

The theoretical results of (\ref{F:n}-\ref{Sxyz}) for 
the strength of fluctuations in the RNA cascade are independent of the details of the enzyme biogenesis.
However, for simplicity we limit the enzyme biogenesis in numerical simulations to the following birth-death processes
\begin{equation}
\def\arraystretch{0.5}  %  reduce row spacing
\begin{array}{ccccc}
\varnothing & \Orightarrow{\kappa A} &  \alpha  & \Orightarrow{\kappa} &  \varnothing, \\
\varnothing & \Orightarrow{\kappa B} &  \beta  & \Orightarrow{\kappa} &  \varnothing,
\end{array}  \label{RR:e}
\end{equation}
where $A$, $B$ are the enzymatic expression levels at steady-state and
$1/\kappa$ governs the time scale during which enzyme expression levels remain constant.
In Figure~\ref{fig:xyz} we numerically test the predictions of (\ref{F:n}) based on the enzyme biogeneses of (\ref{RR:e}).
We show the linear dependence on $k_x$ of the Fano factors computed from $10^5$ simulations of (\ref{RR:RNA}-\ref{RR:e}) using the Gillespie algorithm~\cite{Gillespie1977}
with $x$, $y$ and $z$ collected after the system reached a steady-state.
The kinetic parameters used in the simulations correspond to the siRNA biogenesis of \textit{Salmonella}~\cite{Overgaard2009}.
While prokaryotes and eukaryotes have different sRNA biogeneses, differences are quantitative rather than qualitative.
Replacing the microprocessor/Dicer activity with the Cascade/Cas activity~\cite{Vanderoost2009}, then (\ref{R:miRNA-inter},\ref{RR:RNA}) are still applicable.
Additionally, we numerically simulated (\ref{RR:RNA}-\ref{RR:e}) in the range of parameters associated with mammalian mRNA expression~\cite{Schwanhausser2011}, 
and miRNA activity in mammals~\cite{Selbach2008,Guo2010,Baccarini2011} and obtained similar results in each case (data not shown).

In the main plot of Figure~\ref{fig:xyz}, Fano factors order according
to $F_z < F_x < F_y$ for different values of the  transcription rate
and the speed of fluctuations in enzyme copy-numbers ($\kappa \geq 0.1/$h).
It seems that the two-step RNA processing cascade (\ref{RR:RNA}) amplifies fluctuations in $y$ but filters fluctuations in $z$
under certain conditions and to a certain extent.
Fluctuations in the intermediate product $y$ are expected to be stronger than fluctuations in the transcript $x$, 
because both the birth and death rates of $y$ vary with time, whereas only the death rate of $x$ varies with time. 
However, the birth rate of the mature product $z$ follows closely the fluctuations that $y$ undergoes. 
One would expect the possibility $F_z > F_y$ to arise as well.
These observations are reflected in (\ref{Sxyz}), where $S_x$ depends only on $\alpha$, 
whereas both $S_y$, $S_z$ depend on $\alpha$ and $\beta$.
In fact, $S_y$ and $S_z$ differ only by the term $S_z/S_y = k_z\beta/d_z$.
This term, along with the speed of enzyme fluctuations $\kappa$, determines the differences observed among $F_y$ and $F_z$ in Figure~\ref{fig:xyz}.
For given $\kappa$, if $k_z\beta/d_z \gg 1$ then fluctuations in the ancestor substrates are amplified in the mature RNA product, 
whereas noise filtering takes place if $k_z\beta/d_z \ll 1$.
During a given time interval and when $k_z \beta \gg d_z$, there are on average many more $z$-production than $z$-degradation events
and noise amplification occurs. 
On the other hand, when $k_z \beta \ll d_z$, the $z$-degradation events overwhelm the $z$-production events resulting into noise filtering.
This effect shows up in the inset of Figure~\ref{fig:xyz}, which shows results for $\kappa = 0.1$/h and a twice as stable mature RNA product ($d_z=0.5$/h) compared to the main plot.
In this case, we observe $F_x \simeq  F_y < F_z$:
the RNA cascade amplifies the strength of fluctuations in the mature RNA product (blue lines). 
However, when the processing rate is reduced $k_z \beta \ll d_z$, the situation reverses again (red lines) and $F_x \simeq F_y > F_z$:
the RNA cascade buffers the noise in the expression level of the mature RNA product as is also observed in the main plot of the figure.

The majority of mature RNA products are expected to be more stable than their precursors.
For example, mature miRNAs when loaded to Argonaute proteins are stabilized, 
and in certain cases in vitro half-lives become longer than a day~\cite{Baccarini2011,Gantier2011}.
Mature transcripts are also protected by 5'-capping and polyadenylation.
Thus for most cases we expect $d_z<\{d_x,d_y\}$, leading to the
amplification of fluctuations through the processing cascade.
On the other hand, in cases where the processing rate per precursor molecule is lower than $d_z$, 
or the mature RNA product is unusually unstable, 
the RNA cascade will operate in the reverse regime and reduce the strength of fluctuations in the expression level of the mature RNA product.

The key result of Eq.~(\ref{F:n}) is that when RNA-processing enzyme copy-numbers fluctuate slower than their substrates,
a gene's transcription rate affects the copy-number fluctuations of the mature product, whether it is protein or small RNA.
As we are going to show, this remains true also when transcription takes place in bursts due to fluctuations in the promoter state~\cite{Swain2002,Paulsson2005}.
As the copy numbers of processing enzymes affect the processing of the products of several genes, 
fluctuations of processing enzymes can be viewed as a contribution to ``extrinsic stochasticity" of gene expression~\cite{Swain2002,Paulsson2005}.
Below, we discuss how to pick up this signature experimentally.

\subsection{Dynamics of production events in the RNA cascade}
\label{dynamics}

We have shown that in the absence of enzymatic copy-number fluctuations any product of the cascade (\ref{R:RNA})
is Poisson-distributed at steady-state.
We now show that the dynamics of $z$ production in (\ref{R:RNA}), and thus correlations to any order, are identical to the dynamics and correlations  
of a homogeneous Poisson process of rate $\widetilde{k}_z$.
For simplicity, let us investigate a variation of (\ref{R:RNA}) consisting only of a single processing step
\begin{equation}
\def\arraystretch{0.5}  %  reduce row spacing
\begin{array}{ccccccc}
\varnothing & \begin{array}{c} \Orightarrow{k_x} \\ \Uleftarrow{d_x} \end{array} & x & \Orightarrow{k_y} & y & \Orightarrow{d_y} & \varnothing.
\end{array}\label{r:yz}
\end{equation}
The addition of more processing steps is straightforward to handle.
Since enzyme copy-numbers in (\ref{r:yz}) are fixed, $x$ and $y$ are Poisson-distributed at steady-state.
Additionally, the dynamics of $x$ is a birth-death process
$\varnothing \Orightarrow{k_x} x \Orightarrow{k_y+d_x} \varnothing$ with constant birth and death rates.
In other words, the creation of $x$ molecules is a homogeneous Poisson process.
Once an $x$ particle is created, it either decays, or after a time interval it is converted into a $y$ molecule.
For constant processing rates, this time interval is exponentially distributed and the time instants of $y$-creation events become uniformly distributed.
Considering all such $y$-creation events originating from the rest of the $x$ molecules, introduces random shifts into the time intervals of $y$-creation events.
However, the distribution of the time instants of $y$-creation events remains uniform, and the underlying $y$-creation process remains a homogeneous Poisson process
with a rate of $k_y \langle x \rangle$.
Therefore, the statistics of the process $\varnothing \Orightarrow{k_y x(t)} y$ 
becomes identical to the statistics generated by the process $\varnothing \Orightarrow{k_y \langle x \rangle} y$.
The same logic applies if more steps are added in the processing cascade (\ref{r:yz}).
As we discuss below, this implies that the autocorrelation function of any component of the cascade is identical to the autocorrelation function 
of a birth-death process at constant effective rates.

The situation is different when the copy-number of the processing enzyme $\alpha(t)$ fluctuates over time, 
leading to changes in the $x$-processing rate $k_y \alpha(t)$.
All $x$ molecules present at a given time are either processed at a higher or at a lower rate depending on the value of $\alpha(t)$.
According to our discussion so far, within the time interval of the order of $1/\kappa$ where $\alpha(t)$ is constant,
all $y$-creation events follow a homogeneous Poisson process of rate $k_y\alpha(t)\langle x\rangle$.
As $\alpha(t)$ changes however, this rate changes also, resulting to inhomogeneities of size of the order of $1/\kappa$ in the overall distribution of $y$-creation times.
Consequently, the homogeneous Poisson process $\varnothing \Orightarrow{k_y \langle \alpha x \rangle} y$ with uniformly distributed in time $y$-creation events
produces different statistics than the process $\varnothing \Orightarrow{k_y \alpha(t) x(t)} y$.

\subsection{Autocorrelation function of the mature RNA product}

The autocorrelation function $C_z(\tau)= \langle z(t+\tau) z(t) \rangle - \langle z(t+\tau) \rangle \langle z(t) \rangle$ is a measure that identifies temporal correlations 
in the copy-number of the mature RNA product $z$ in the cascades (\ref{R:RNA},\ref{RR:RNA}).
When $\tau\rightarrow 0$, one expects $z(t+\tau)$ to be highly correlated with $z(t)$.
When $\tau \rightarrow \infty$ one expects of them to decorrelate, that is, 
one expects all ``memory" of the value of $z(t)$ to be lost when we resample $z$ at a much later time-point.

\begin{figure}[!htb]
\begin{center}
\includegraphics[totalheight=11.0cm, keepaspectratio=true, viewport=0 0 795 700,clip]{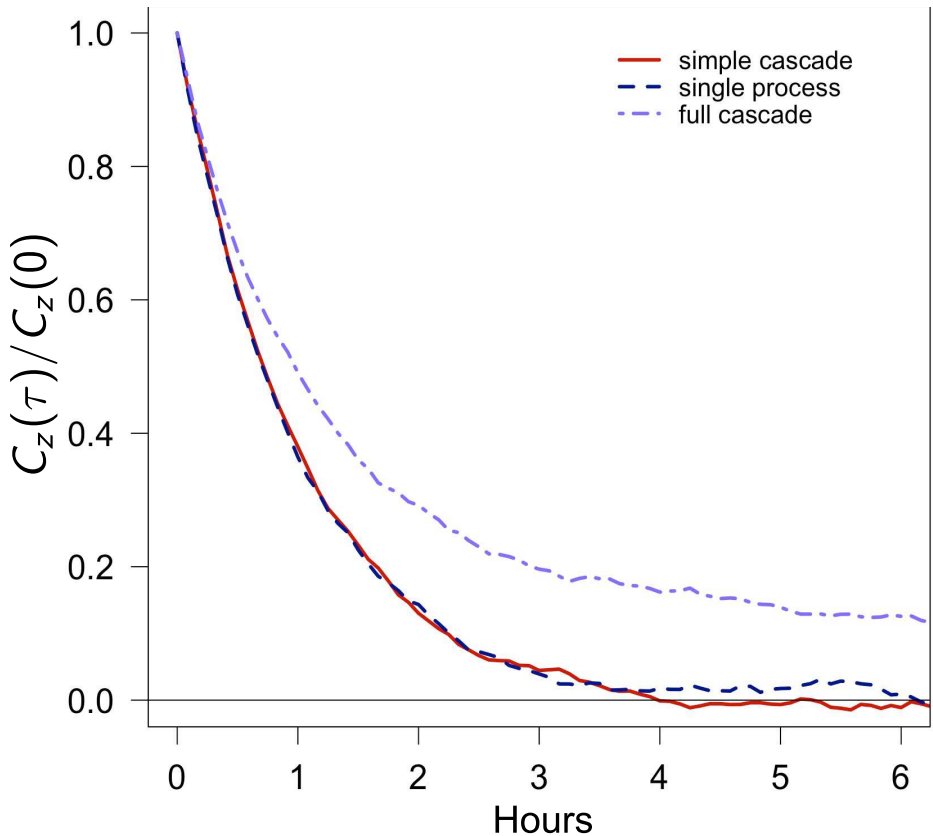}
\end{center}
\caption{\small
{\bf Autocorrelation function of the mature product in the RNA cascade.}
The normalized autocorrelation function of the mature product $z$ is evaluated at steady-state for different time-points 
and ensemble-averaged over $10^4$ realizations of the system.
Identical parameters as in Figure~\ref{fig:xyz} are used. The transcription rate of $x$ is set to $k_x = 20/$min.
We plot $C_z(\tau)/C_z(0)$ for
(i) the simple cascade (\ref{R:RNA}) (red solid line), 
(ii) a single birth-death process with identical $z$-degradation rate as in (i) and $z$-creation rate equal to the average creation rate in (i) (dark-blue dashed line),
and (iii) the full cascade (\ref{RR:RNA}) with fluctuating enzyme numbers ($\kappa = 0.1$/h) and otherwise identical parameters as in (i) (light-blue dash-dotted line).
At constant rates, there is no distinction in the dynamics between (i) and (ii).
In the presence of enzyme fluctuations correlations persist over longer times of the order of $1/\kappa$.
}
\label{fig:cor}
\end{figure}

In Figure~\ref{fig:cor} we plot $C_z(\tau)/C_z(0)$ for three different systems:
(i) the simple multistep cascade (\ref{R:RNA}) with constant reaction rates,
(ii) a single birth-death process of constant rates $\varnothing \Orightarrow{\widetilde{k}_z} z \Orightarrow{d_z} \varnothing$
with $\widetilde{k}_z = k_z \langle y \rangle$ and $\langle y \rangle$ taken from (i) at steady-state, 
and (iii) the full multistep cascade (\ref{RR:RNA}) with a speed of enzyme copy-number fluctuations determined by $\kappa = 0.1$/h.
For any range of $\tau$, there is no significant difference in the autocorrelation function between (\ref{R:RNA}) 
and a single birth-death process of constant rates.
In line with our previous discussion, differences appear when enzymatic fluctuations are included.
In this case, enzyme fluctuations taking place over $1/\kappa$-sized time intervals affect the corresponding substrates.
The memory of this effect across all $z$ is embedded in the autocorrelation function: temporal correlations persist over intervals of the order of $1/\kappa$.

\subsection{Effects due to RNA-processing remain prominent in the presence of transcription bursts}
\label{bursts}

Gene transcription is a process with a significant level of intrinsic
noise already without fluctuations from RNA processing~\cite{Golding2005}.
A strong promoter ensures that a gene is transcribed most of the time, and pauses occur over short intervals only.
A weak promoter on the other hand, results into longer pauses of
transcription leading to strong fluctuations in the copy number of
primary transcripts.
RNA-processing takes place downstream of transcription.
As a result, fluctuations in RNA-processing adds to the fluctuations due to bursty transcription. 

\begin{figure}[!htb]
\begin{center}
\includegraphics[totalheight=11.0cm, keepaspectratio=true, viewport=0 0 1600 1415,clip]{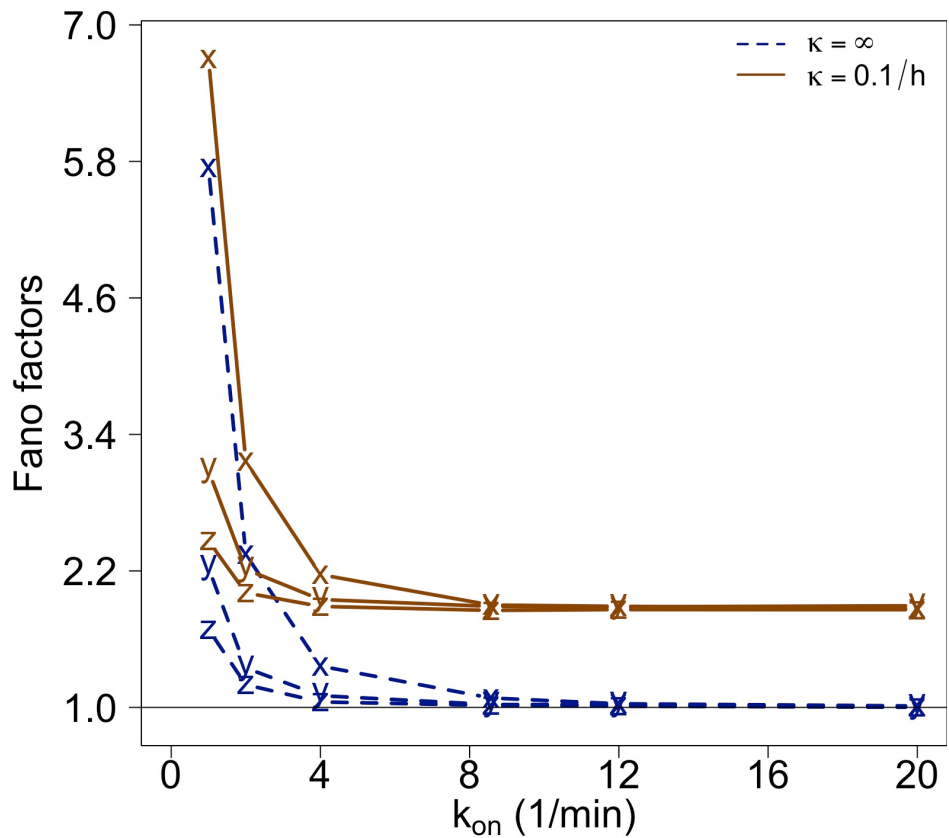}
\end{center}
\caption{\small
{\bf RNA processing with transcription bursts.}
Fano factors of the RNA substrates $x,y,z$ in the chain (\ref{RR:RNA}) 
are plotted as functions of the rate of transcription activation $k_{\mathrm{on}}$.
The rate of transcription inactivation is fixed at $k_{\mathrm{off}}=0.1$/min~\cite{Golding2005}.
When transcription is active the transcription rate is $k_x = 60$/min.
The remainder of the parameters are identical to Figure~\ref{fig:xyz}.
In the absence of enzyme copy-number fluctuations ($\kappa = \infty$,
dashed blue lines) and when $k_{\mathrm{on}} \sim k_{\mathrm{off}}$ 
transcription bursting leads to non-Poissonian statistics.
This deviation from Poisson statistics is further enhanced by fluctuations in RNA-processing 
downstream of transcription ($\kappa = 0.1$/min, solid brown lines).
}
\label{fig:transcription_bursts}
\end{figure}

Figure~\ref{fig:transcription_bursts} shows the Fano factors of the RNA substrates at fixed transcription rate 
as a function of the rate of transcription activation in the absence (dashed blue lines) and in the presence (solid brown lines) of enzymatic fluctuations.
As expected, Fano factors in the absence of enzymatic fluctuations collapse to unity in the limit of strong transcription activation~\cite{Thattai2001},
whereas in the presence of slow enzymatic fluctuations we recover the result of Figure~\ref{fig:xyz}.
Additionally, fluctuations in the number of processed RNA increase as the rate of transcription activation is reduced and transcription becomes more irregular.
The signature of RNA-processing remains prominent throughout the whole range of transcription activation values.
This is despite the conservative assumption in our numerical simulations that enzyme copy-number fluctuations are Poissonian.
If RNA-processing enzymes are produced in bursts as well~\cite{Shahrezaei2008}, 
the effect on RNA-processing would be stronger than what Figures~\ref{fig:xyz},\ref{fig:transcription_bursts} show.

\subsection{Fluctuations in RNA-processing impact on protein biogenesis}

Protein production takes place in bursts due to multiple translation events of mRNA transcripts
even when transcription is constitutive and the mRNA biogenesis is a birth-death process of constant rates~\cite{Thattai2001}.
Introducing RNA enzyme fluctuations in the mRNA biogenesis is expected to render protein production even more noisy.
Here, we investigate within our conservative framework how much
fluctuations in protein production increase due to fluctuations in the copy-numbers of the RNA-processing enzymes.

\begin{figure}[!htb]
\begin{center}
\includegraphics[totalheight=11.0cm, keepaspectratio=true, viewport=0 0 1600 1415,clip]{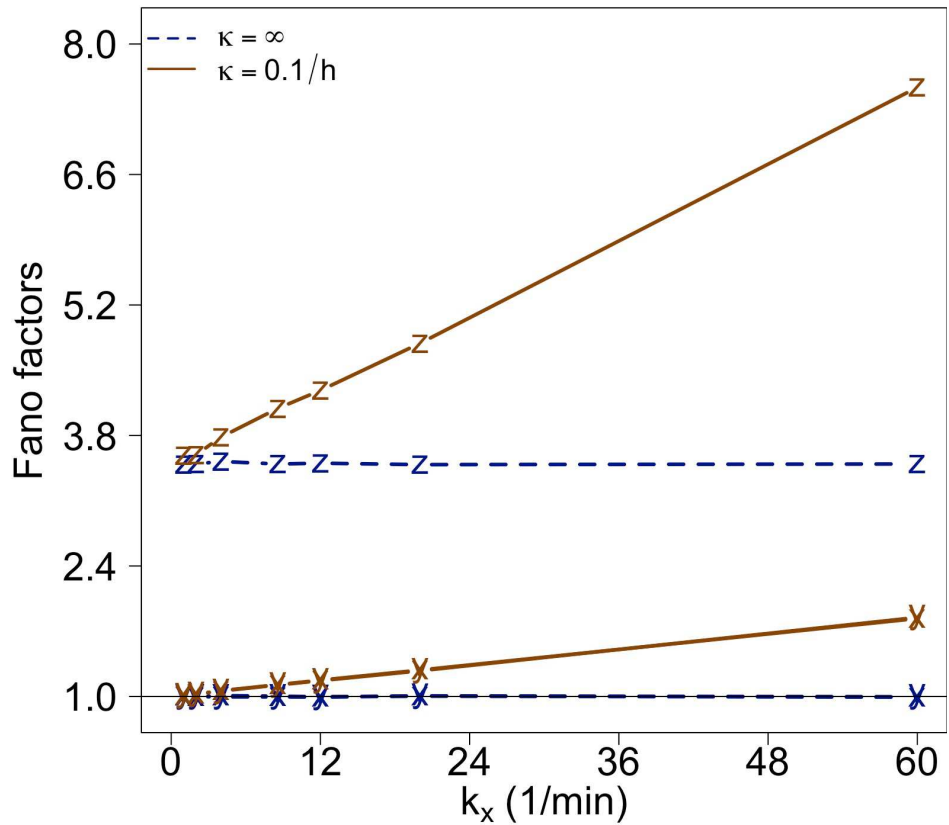}
\end{center}
\caption{\small
{\bf Effect of RNA processing on protein biogenesis.}
Fano factors of pre-mRNA ($x$), mRNA ($y$) and protein ($z$) are plotted as functions of the gene transcription rate $k_x$
in the absence ($\kappa = \infty$, dashed blue lines) and in the presence ($\kappa = 0.1$/min, solid brown lines) of spliceosome copy-number fluctuations.
The translation rate per mRNA molecule is fixed at $k_z = 5$/h and the rest of parameters are identical to Figure~\ref{fig:xyz}.
In the absence of spliceosome fluctuations the statistics of $x$, $y$
is Poissonian, but bursts of translation lead to non-Poissonian
statistics of $z$ 
with a Fano factor $F_z$ independent of $k_x$~\cite{Thattai2001}.
When spliceosome copy-number fluctuations are included, fluctuations in protein production increase linearly with $k_x$.
Our results are conservative since fluctuations in ribosome copy-numbers are ignored, and spliceosome fluctuations are taken to be Poissonian.
}
\label{fig:translation_bursts}
\end{figure}

Figure~\ref{fig:translation_bursts} shows the Fano factors of pre-mRNA ($x$), mRNA ($y$) and protein ($z$) as functions of the gene transcription rate
in the absence (dashed blue lines) and presence (solid brown lines) of spliceosome copy-number fluctuations.
Fano factors of the RNA substrates collapse to unity as expected in the absence of enzymatic fluctuations, 
and the protein Fano factor becomes $1 + (k_z/d_y) / (1+d_z/d_y) = 3.5$ and independent of the transcription rate~\cite{Thattai2001}.
In the presence of RNA-processing enzyme copy-number fluctuations however, we find that the protein biogenesis is significantly affected
and protein noise becomes linearly dependent now on the transcription rate.
We stress again that these results are conservative because RNA-processing enzyme fluctuations are modelled as Poissonian.
Additionally, copy-number fluctuations of the ribosomal units are not included in the simulation.
Relaxing any of these two conditions would only amplify the strength of the effect shown in Figure~\ref{fig:translation_bursts}.

\section{Conclusion}

RNA biogenesis is a multi-step cascading process for protein-encoding transcripts and sRNAs alike.
The output of one process becomes the input of the next one until the final mature product is reached.
We showed that when reactions in the RNA cascades occur at constant rates, 
then the mature products undergo single-step birth-death biogeneses
of constant effective rates and are Poisson-distributed at steady-state.
This simple picture breaks down when there are fluctuations in the copy-numbers of enzymes that mediate RNA-processing.
Enzymatic fluctuations induce fluctuations in the processing rates per corresponding substrate molecule.
We showed that Fano factors of the RNA cascade's products increase linearly with the transcription rate, 
irrespective of the form of the steady-state distribution of the copy-numbers of enzymes participating in the processing steps.
In numerical simulations that include transcriptional and
translational bursting we find that this effect remains detectable in
the presence of other sources of fluctuations, especially
for the case of protein biogenesis.

Post-transcriptional regulation is a significant part of sRNA biogenesis.
Mature sRNAs are recruited by Argonaute proteins in order to form the RNA-induced silencing complexes (RISCs)~\cite{Krol2010}.
RISCs are the units that mediate PTR, but also ensure the stability of sRNAs, rendering them important elements of the sRNA biogenesis.
Additionally, early sRNA processing steps can be affected in numerous ways via feedback regulation.
One example is miR162 regulating DCL1; further examples include miR168 regulating the Argonaute protein AGO1, and miR403 regulating AGO2~\cite{Allen2005}.
Furthermore, there is abundant evidence of feedback regulation to sRNA genes by their transcription factor targets~\cite{Lutter2010}.
All of this dynamics remained outside the scope of our analysis, as it involves PTR rather than the sRNA biogenesis.
However, if feedback regulates the transcription of sRNA only, then the conclusions of our analysis are still applicable.
If enzyme fluctuations are negligible, one can replace the sRNA cascade with a single birth-death process and incorporate the feedback regulation into the sRNA birth process. 
If on the other hand, like the case of miR162 and DCL1, feedback regulation takes place in sRNA processing, 
then clearly the processing details matter and need to be included.
This can be the topic of a future study of feedback dynamics within the RNA-processing cascade.
Finally, many RNA subclasses are not mentioned in this work. 
For example, piwi-interacting RNAs, small nucleolar RNAs, or small nuclear RNAs, all have distinct biogeneses and functionalities~\cite{Guil2012}.
If these RNA subclasses follow processing chains like (\ref{R:RNA}), 
or the more general (\ref{RR:RNA}), then our results are applicable for them also.

Single nucleotide polymorphisms affect the processing rate of precursors of sRNAs and impact on the expression level of the mature products~\cite{Todesco2012}.
Naturally, polymorphisms affecting transcription in sRNA induces also
variation across species in the production of mature sRNAs.
However, based on our analysis of the steady-state expression levels in the sRNA biogenesis, 
we predict that the effect of transcriptional variation can be enhanced, reduced, or even reversed by the presence of variation in sRNA processing.
This result shows that the functional consequences of modification in pre-miRNA expression and processing rates are not intuitive and 
should depend on the evolutionary forces acting on the final miRNA product.
In situations where a miRNA is under positive selection to be expressed at higher steady state levels, 
increased processing might provide a greater advantage than an increase in the rate of transcription,
even if the mutation occurs in a genetic background with a weak transcription rate.
Conversely, in the face of evolutionary constraints for maintaining steady state levels, 
one might observe compensatory evolution between mutations altering processing and transcription rates.
Within population, this might lead to stronger linkage disequilibrium between the upstream regulatory region and the region encoding the miRNA precursor.
To date, functional changes in miRNA encoding genes during evolution were seldom examined,
but the few studies often showed that both miRNA transcription and processing rates vary~\cite{Ng2011,Wangetal2011}.
Evaluating the extent of compensatory evolution requires experimental systems where the degradation products of miRNA processing can be
quantified so that transcription and processing rates can be estimated. 
Deep sequencing technologies are now making such estimates possible. 

Our analysis showed that the RNA cascade amplifies the noise
in the expression level of a mature RNA product provided the mature
product is more stable than its precursors (as is generally the case).
However, we showed it is also possible for the RNA cascade to reduce the noise in the expression level of unusually unstable mature RNA products,
or of those mature products whose precursors are processed at low rates.

The simplest experimental test of our results would involve a Tet-inducible promoter that drives the expression of a fluorescent protein,
whose transcript contains at least one intron in order to ensure that the spliceosome will interact with it.
Spliceosome copy-number fluctuations can be induced indirectly by specific inhibitors~\cite{Kuhn2009,Effenberger2013}.
Induced expression of the fluorescent protein under slow spliceosome copy-number fluctuations will result in much broader protein distributions
due to the broadening of the mRNA distributions according to Eq.~(\ref{F:n}).
Furthermore, we expect a stronger expression of the fluorescent protein under fixed spliceosome inhibitor concentrations to result also in broader protein
distributions with Fano factors linearly depending on the tetracycline concentration as we showed in Figure~\ref{fig:translation_bursts}.

In summary, we investigated the RNA processing cascade and found universal characteristics in the steady-state dynamics for different RNA species.
If processing steps take place at constant rates, then the mature RNA biogenesis can be modeled at steady-state as a single birth-death process of constant rates.
Variation in the processing rates induces additional fluctuations to the RNA cascade and the single birth-death picture breaks down.
Finally, we showed that polymorphisms in the processing rates can act synergistically or antagonistically to polymorphisms in the transcription rates of RNA.
Our work offers a framework to better understand the dynamics of RNA biogenesis and a fortiori of RNA evolution.

\section{Acknowledgments}

This work was supported by Deutsche Forschungsgemeinschaft (DFG) grant
SFB 680, by the BMBF-SysMO2 grant and by SyBaCol.

\section{References}

\bibliography{references}
\bibliographystyle{unsrt}

\end{document}